\documentclass[12pt,twocolumn]{iopart}

\usepackage{iopams} 
\usepackage{graphicx}

\begin{document}

\title[$\beta$-$\mathrm{Ga_2O_3}$ anisotropy in thermal properties]{Temperature-dependent thermal conductivity and diffusivity of a Mg-doped insulating $\beta$-$\mathrm{Ga_2O_3}$ single crystal along [100], [010] and [001]}

\author{M. Handwerg$^{1,2}$, R. Mitdank$^{1}$, Z. Galazka$^{3}$, S. F. Fischer$^{1}$}
\address{1. AG Neue Materialien, Humboldt-Universit{\"a}t zu Berlin, Institut f{\"u}r Physik, Newtonstrasse 15, 12489 Berlin, Germany} \address{2. Helmholtz-Zentrum Berlin f{\"u}r Materialien und Energie GmbH, Hahn-Meitner-Platz 1, 14109 Berlin, Germany} \address{3. Leibniz Institute for Crystal Growth, Max-Born-Strasse 2, 12489 Berlin, Germany}
\ead{handwerg@physik.hu-berlin.de}

\begin{abstract}

The monoclinic crystal structure of $\beta$-$\mathrm{Ga_2O_3}$ leads to significant anisotropy of the thermal properties. 
The 2$\omega$-method is used to measure the thermal diffusivity $D$ in [010] and [001] direction respectively and to determine the thermal conductivity values $\lambda$ of the [100], [010] and [001] direction from the same insulating Mg-doped $\beta$-$\mathrm{Ga_2O_3}$ single crystal.
We detect a temperature independent anisotropy factor of both the thermal diffusivity and conductivity values of $D_{[010]}/D_{[001]}=\lambda_{[010]}/\lambda_{[001]}=1.4\pm 0.1$.
The temperature-dependence is in accord with phonon-phonon-Umklapp scattering processes from 300 K down to 150 K. Below 150 K point-defect-scattering lowers the estimated phonon-phonon-Umklapp-scattering values.
\end{abstract}

\maketitle

\section{Introduction}

Semiconducting transparent oxides like $\beta$-$\mathrm{Ga_2O_3}$ provide a huge potential for future use in high-power electronic applications \cite{introhighpower} and opto-electronics, like thin-film electroluminescent displays \cite{transdis} or transparent field effect transistors \cite{transFET}.
Monoclinic $\beta$-$\mathrm{Ga_2O_3}$ can be grown from the melt and it is the most stable phase in a large temperature range from 4.2~K to 1725~K out of  five different modifications ($\alpha$ - $\epsilon$).
Recently, investigations of optical, electrical \cite{mitdank,introbulk} and thermal properties~\cite{handwerg,Galazka2} have been carried out, but there are still unsolved issues. 
In particular, the thermal properties are yet to be investigated.

Previous measurements on Czochralski grown $\beta$-$\mathrm{Ga_2O_3}$ crystals were carried out for the temperature-dependent thermal conductivity in [100]-direction from room temperature down to 25~K, \cite{handwerg} and for the [010] direction from room-temperature up to 1200~K \cite{Galazka2}.
The room temperature thermal conductivity was found to be anisotropic: $\lambda_{[100]}=(13\pm 1)\mathrm{Wm^{-1}K^{-1}}$, \cite{handwerg} $\lambda_{[010]}=21~\mathrm{Wm^{-1}K^{-1}}$ \cite{Galazka2}.
Additionally, recent investigations of the thermal conductivity of Sn-doped edge-defined film fed grown (EFG) $\beta$-$\mathrm{Ga_2O_3}$ crystals were carried out for all directions from different samples using the time-domain thermoreflectance method \cite{Guo} with room temperature thermal conductivity values of $\lambda_{[100]}\approx (10.9\pm 1.0)\mathrm{Wm^{-1}K^{-1}}$, $\lambda_{[010]}=(27.0\pm 2.0)\mathrm{Wm^{-1}K^{-1}}$ and $\lambda_{[001]}\approx 15\mathrm{Wm^{-1}K^{-1}}$. 
A theoretical investigation from Santia \textit{et al.} \cite{Lambdatheo} calulated thermal conductivity values from first principles with $\lambda_\mathrm{theo,[100]}=16~\mathrm{Wm^{-1}K^{-1}}$, $\lambda_\mathrm{theo,[010]}=22~\mathrm{Wm^{-1}K^{-1}}$ and $\lambda_\mathrm{theo,[001]}=21~\mathrm{Wm^{-1}K^{-1}}$ at 300~K.

This work provides temperature-dependent values of the thermal diffusivity in [010] and [001] direction as well as thermal conductivity in [100], [010] and [001] direction of Mg-doped insulating monoclinic $\beta$-$\mathrm{Ga_2O_3}$ bulk crystals. We implemented the $2\omega$ method in order to determine the thermal diffusivity along different crystal axes using current heating techniques. The complete data set was determined at the same sample. The method proves powerful in order to measure the thermal properties of single crystals along arbitrary directions.
The measurements were performed in a temperature range from 64~K to 300~K.
By applying the electrical line heater 2$\omega$-method we obtain anisotropic values from the very same crystal.
Only recently, this 2$\omega$-method was presented by Ramu \textit{et al.}\cite{Ramu1} to determine the anisotrop thermal properties by using the transport of temperature oscillation between two parallel metal lines. Here, this method is first applied to $\beta$-$\mathrm{Ga_2O_3}$. We demonstrate it to be more powerful than the 3$\omega$-method because of its accurate determination of the thermal diffusivity and conductivity for specific directions.
The doping with magnesium leads to electrical insulation \cite{Galazka2}. This allows us to electrically separate the line heaters from the investigated crystal. 
From the temperature dependence on the thermal properties we obtain insight into the phonon scattering processes. 

\section{Materials and Method}

\begin{figure}[h]
\includegraphics[width=0.7\columnwidth,keepaspectratio]{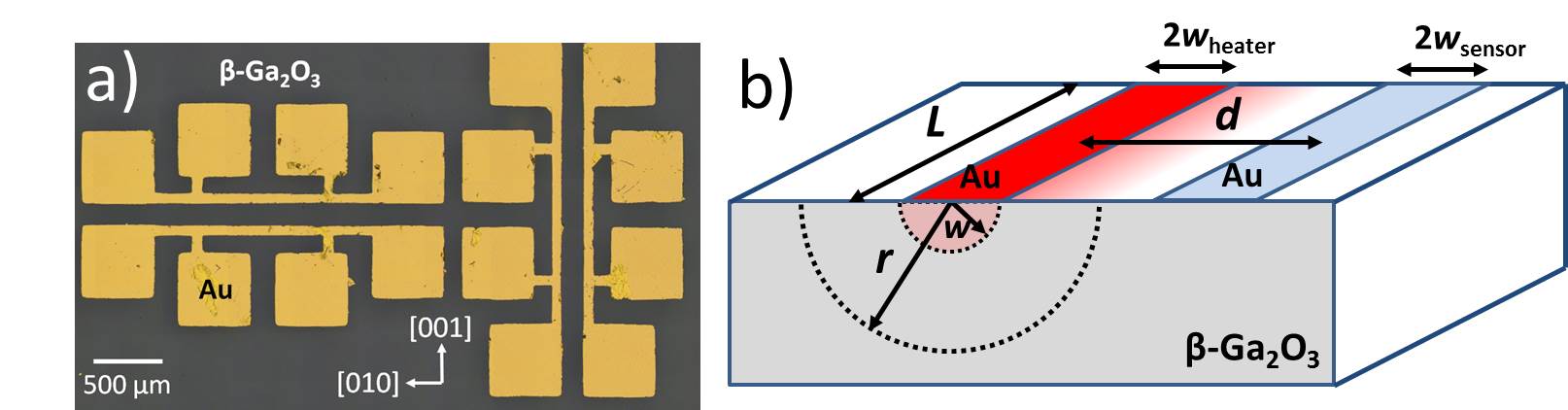}
\caption{a) An arrangement of two line heater pairs on top of the Mg-doped $\beta$-$\mathrm{Ga_2O_3}$-crystal. b) The anisotropic thermal conductivity measurement setup to measure the temperature flow between two metal lines (Au) through the $\beta$-$\mathrm{Ga_2O_3}$-crystal with a thickness of $t=0.5~\mathrm{mm}$. The half metal line width $w=35~\mathrm{\mu m}$, distance of the inner voltage contacts $L=1~\mathrm{mm}$ and the distance between heater and sensor $d=240~\mathrm{\mu m}$ are determined with optical microscopy.}
\label{microbild}
\end{figure}

Our sample specific measurement setup consists of two elements: The Mg-doped insulating bulk $\beta$-$\mathrm{Ga_2O_3}$ single crystal and two metallic heater lines, which are deposited on top of it. These lines can serve as heater and temperature sensors individually.

The crystal was grown from melt by the Czochralski method with use of an iridium crucible and a dynamic, self-adjusting growth atmosphere to minimize decomposition of $\mathrm{Ga_2O_3}$ and oxidation of the iridium crucible. 
The crystal is grown along the [010]-direction with a diameter of 22~mm or 50~mm.
From the bulk crystals epi-polished (100)-oriented wafers were prepared.
A detailed description of the growth process can be found in references \cite{Galazka2,Galazka1}.
With an Mg-concentration of about $n_\mathrm{Mg}=11~\mathrm{wt.ppm.}$ the sample becomes electrically insulating. 
This is due to the compensation of oxygen vacancies (donors) with Mg \cite{Galazka1}. The very low concentration of Mg will only affect the lattice thermal properties at low temperatures. Our previous measurements \cite{handwerg} have shown, that there is no difference in the thermal conductivity of doped or undoped $\beta$-$\mathrm{Ga_2O_3}$ from 150 K to~300 K.
 
Metal-lines were patterned by laser lithography (positive resist AZECI 3027) and an Au-film (50~nm) was sputtered.
Subsequently, a lift-off was performed with acetone in an ultrasonic bath.
Figure \ref{microbild} a) shows two typical heater line structures.

The metal heater lines are operated by electrical joule heating using an AC current. 
In order to determine the generated temperature oscillations $\Delta T$, the temperature dependent resistance $R=R_0+R_0\alpha\Delta T$ of the metal lines is measured. Here, $R_0$ depicts the resistance at the bath temperature $T_0$. 
The temperature coefficient $\alpha$ is determined by $\alpha=(1/R_0)\cdot \partial R/\partial T$. The temperature dependence of the resistance of metals is given by the Bloch-Gr{\"u}neisen-Law~\cite{tritt} 
\begin{equation}
R(T)-R(4.2~\mathrm{K})\propto \left(\frac{T}{\theta_\mathrm{D}}\right)^5\int_0^{\theta_\mathrm{D}/T}\frac{x^5 }{(e^x-1)(1-e^{-x})}\mathrm{d}x.
\label{BG}
\end{equation}
\begin{figure}[h]
\center
\includegraphics[width=0.7\columnwidth,keepaspectratio]{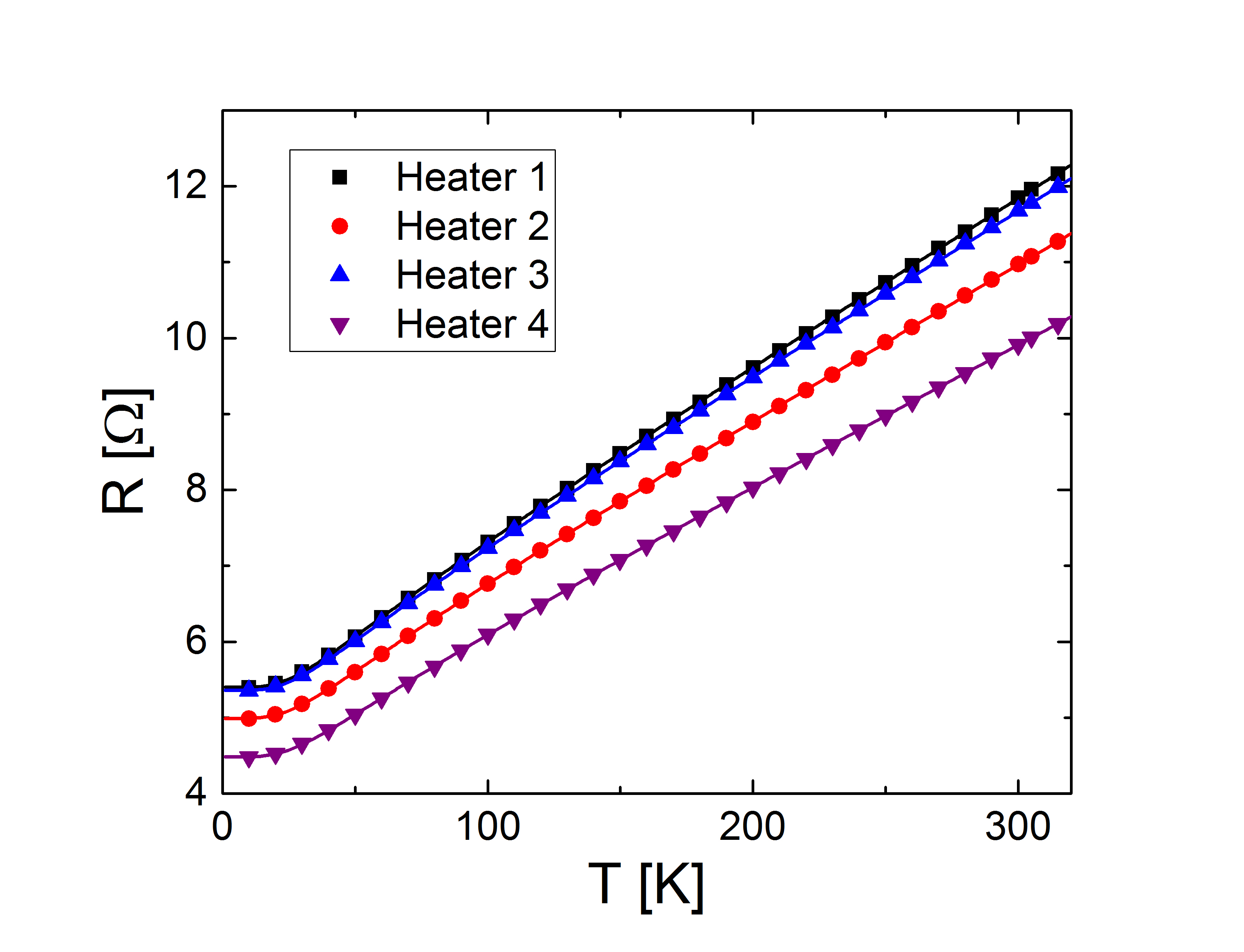}
\caption{The temperature dependent resistance of the four heater lines approximated with equation \ref{BG}, the so called Bloch-Gr{\"u}neisen-law. The temperature coefficient $\alpha(T)$ for both heater lines is calculated from the numerical derivation of the fit curve.}
\label{widerstand}
\end{figure}
As shown in figure \ref{widerstand} the measured heater resistance $R$ is in accord with equation \ref{BG}. From the numerical differentiation of the fitted values from equation \ref{BG} we determine $\alpha(T)$ in the full range of temperatures of our investigations (64 K to 300 K).\\
Joule heating is achieved with an alternating power $P=U\cdot I$ with the oscillating voltage $U=U_0\cdot \cos(\omega t)$ and an oscillating current $I=I_0\cdot \cos(\omega t)$. This leads to an oscillating temperature change $\Delta T\propto P=(U_0 I_0/2)\cdot (1+cos(2\omega t))$ (see ref\cite{Cahill2}) with the angular frequency $\omega=2\pi f$ and $f$ denotes the measurement frequency.  
These temperature oscillations are transported within the crystal according to its thermal properties. 
To detect temperature oscillations in a metallic sensor line in a distance $d$ parallel to the heater with a direct current $I_\mathrm{0,DC}$, a resistance $R_\mathrm{S}$ and a temperature coefficient $\alpha_\mathrm{S}$, one needs to detect the correlated resistance oscillations $R_\mathrm{S}(2\omega)\propto\alpha\Delta T_\mathrm{S} (2\omega)$ and therefore the measureable voltage oscillations of two times the heater frequency $U_\mathrm{2\omega}(2\omega)=I_\mathrm{S,DC}\cdot R_\mathrm{S}(2\omega)$. 
Here, voltage oscillations are detected as a function of the transported temperature oscillations $\Delta T_\mathrm{S}$ by

\begin{equation}
U_\mathrm{2\omega}=\frac{\alpha_\mathrm{S} I_\mathrm{S,DC} R_\mathrm{S} \sqrt{2}}{2}\Delta T_\mathrm{S}\quad. 
\label{2wmeas}
\end{equation}

\begin{figure}[h]
\includegraphics[width=0.7\columnwidth,keepaspectratio]{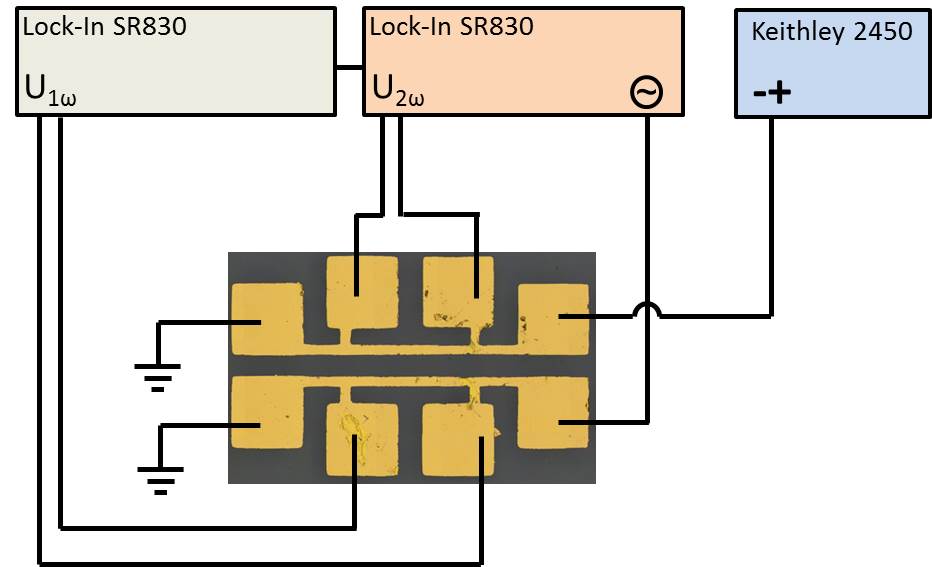}
\caption{The 2$\omega$-method is based on two seperate electrical circuits, one heaterline with dominating alternating heating current and the sensorline with direct current and heat induced second harmonic voltage $U_{2\omega}$. }
\label{schema}
\end{figure}

A Lock-In amplifier (SR830) is used to detect these higher harmonic voltages. A schematic and description is given in figure \ref{schema}.\\
From the frequency dependence of the voltage signal $U_{2\omega}$ the thermal properties can be determined as described below.

\section{Results}

\begin{figure}[h]
\includegraphics[width=0.7\columnwidth,keepaspectratio]{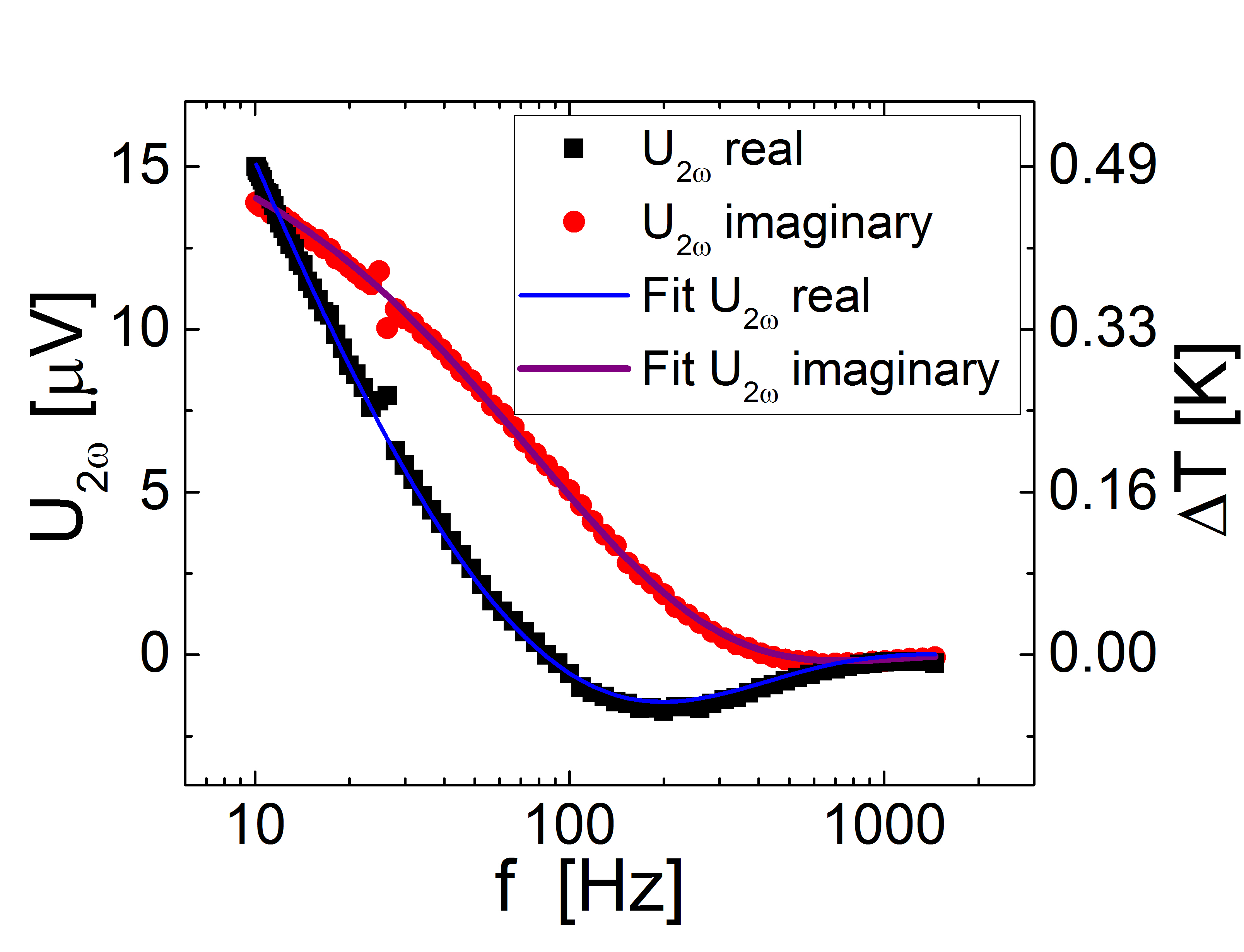}
\caption{The 2$\omega$-voltage oscillations were measured with displaced heater and sensor line perpendicular to the [001]-direction. The approximation was performed using equation \ref{2ww}. The plot was by way of example taken at 250~K.}
\label{U3wundU2w}
\end{figure}

The determination of the thermal diffusivity and conductivity by the 2$\omega$-method is depicted as follows.
The thermal conductivity $\lambda$, thermal diffusivity $D$ and specific heat capacity $C_V$ are related to each other by the equation
\begin{equation}
\lambda = D\cdot C_V \quad .
\label{lamm}
\end{equation}
The thermal conduction differential equation
\begin{equation}
\frac{\partial^2 \Delta T(r,t)}{\partial r^2}+\frac{1}{D} \frac{\partial \Delta T(r,t)}{\partial t}=0
\end{equation}
can be solved \cite{Cahill1} for an oscillating temperature $\Delta T$ to
\begin{equation}
\Delta T(r)=\frac{P}{\pi L_\mathrm{heater} \lambda}\cdot K_0(qr)
\label{ori}
\end{equation}
with $P/L_\mathrm{heater}$ the power per unit length of the heater, $K_0$ the zero-order Bessel function of the second kind and $q$ represents the frequency dependent inverse thermal penetration depth
$q(f)=\sqrt{4\pi f/D}$.
In order to determine the thermal diffusivity $D$ along the crystal axes [010] and [001], equation \ref{ori} has to be utilized to determine the heat flow between the heater and the sensor lines by averaging their widths \cite{ramu2}
\begin{equation}
\Delta T=\frac{P}{\pi L \bar{\lambda}} \frac{1}{2w_\mathrm{heater}} \int^{w_\mathrm{heater}}_{-w_\mathrm{heater}}\frac{1}{2w_\mathrm{sensor}} \quad \times\\ \int^{w_\mathrm{sensor}}_{-w_\mathrm{sensor}}K_0\left(\left( q_\mathrm{[x]}\cdot(d+o-p\right)\right)\mathrm{d}o\mathrm{d}p\quad .
 \label{2ww}
\end{equation}
Here, we denote the inverse thermal penetration depth as $q_{[x]}=\sqrt{4\pi f/D_{[x]}}$, the half heater line width as $w_\mathrm{heater}$, the half sensor line width as $w_\mathrm{sensor}$, the heating power as $P$, the length of the heater line as $L_\mathrm{heater}$ and the average distance between the heater and the sensor line as $d$. The integration parameters $o$ and $p$ depict the distance along the line heater widths.
The effective thermal conductivity is $\bar{\lambda}=\sqrt{\lambda_\mathrm{[x]}\cdot\lambda_\mathrm{[y]}}$ (see ref. \cite{Ramu1}).
An exemplary measurement of $\Delta T(f) \propto U_{2_\omega}(f)$ is shown in figure \ref{U3wundU2w} together with a fit using equation \ref{2ww} by varying the thermal diffusivity $D_{[x]}$ and the thermal conductivity $\bar{\lambda}$. The integrals were solved numerically.
The axis $[x]$ denotes the direction from the heater to the sensor line and $[y]$ the direction perpendicular to the surface plane, here [100].

Here, two heater and sensor lines are placed along the [001] and [010] direction. Therefore, the thermal diffusivity values $D_{[010]}$ and $D_{[001]}$ are directly obtained (figure \ref{D}), as well as the effective thermal conductivity values $\sqrt{\lambda_{[100]}\cdot\lambda_{[010]}}$ and $\sqrt{\lambda_{[100]}\cdot\lambda_{[001]}}$ shown in figure \ref{l}.

With the knowledge of the isotropic specific heat capacity $C_V$ one can determine the thermal conductivity values for all three primary axes as shown in table \ref{tabletop}. 

\section{Discussion}
\begin{figure}[h]
\center
\includegraphics[width=0.7\columnwidth,keepaspectratio]{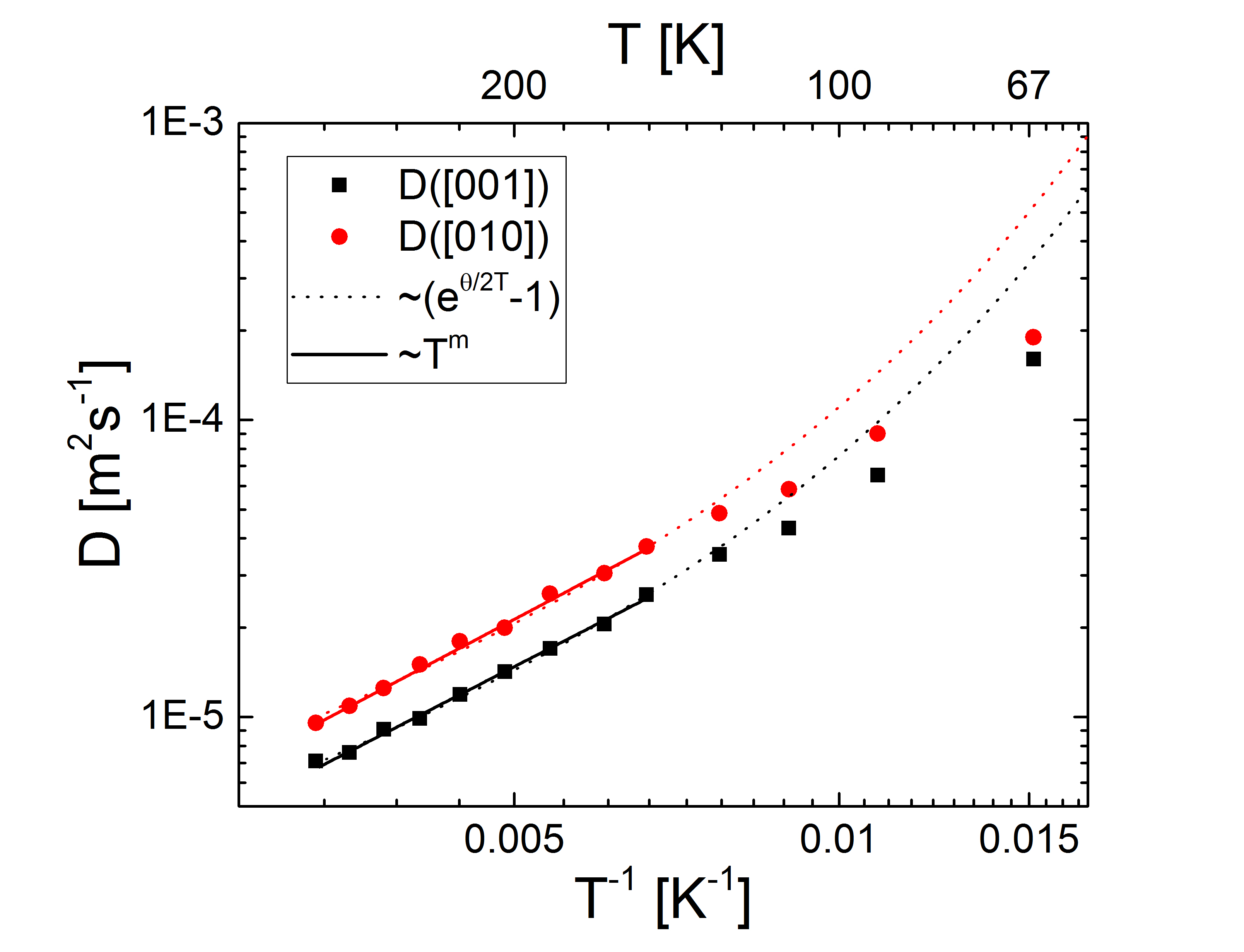}
\caption{Temperature-dependent thermal diffusivity of insulating $\beta$-$\mathrm{Ga_2O_3}$ on a logarithmic plot vs. the inverse temperature. The approximaton $D(T)=a\cdot T^m$ (solid line) reveals the same slope $m$ for both directions. The measurements agree with phonon-phonon-Umklapp scattering (equation \ref{LAM}, dotted line) in the high temperature regime from 150~K to 300~K. Below 150~K, point defect scattering presumably lowers the diffusivity values.}
\label{D}
\end{figure} 
From the $2\omega$-measurements we obtained directly the temperature dependent thermal diffusivities in [010]- and [001]-direction as shown in figure \ref{D}.
We observe higher thermal diffusivity values in [010]-direction then in [010]-direction.
For high temperatures from 150 K to 300 K we observe a linear dependence in logarithmic scale for both directions with $D(T)=a\cdot T^m$ (solid lines). 
Both linear approximations have the same slope of $m=1.9\pm 0.1$ which means the anisotropy factor is $D_{[010]}/D_{[001]}=1.4\pm 0.1$ and temperature independent.
The temperature dependence of the diffusivity is determined by the temperature dependence of the phonon mean free path $\Lambda_\mathrm{ph}$ and the temperature independent velocity of sound $v_\mathrm{s}$ as 
\begin{equation}
D(T)=\frac{1}{3}\Lambda_\mathrm{ph}(T)\cdot v_\mathrm{s}\quad.
\end{equation}
The temperature dependence of the phonon mean free path on behalf of phonon-phonon interaction with Umklapp-scattering would be \cite{hunklinger} 
\begin{equation}
\Lambda \propto \left(e^{\theta_\mathrm{D}/2T}-1\right) 
\label{LAM}
\end{equation}
with $\theta_\mathrm{T}$ as Debye-temperature.
The approximation of the diffusivity values with equation \ref{LAM} (dotted lines) fits well in the temperature range from 150 K to 300 K.
Below 150 K point-defect-scattering lowers the estimated thermal diffusivity presumably by inhibiting the phonon mean free path with an additional scattering effect.
The Mg-concentration dominates all other defects and is most likely the cause of point-defect-scattering. 

\begin{figure}
\center
\includegraphics[width=0.7\columnwidth,keepaspectratio]{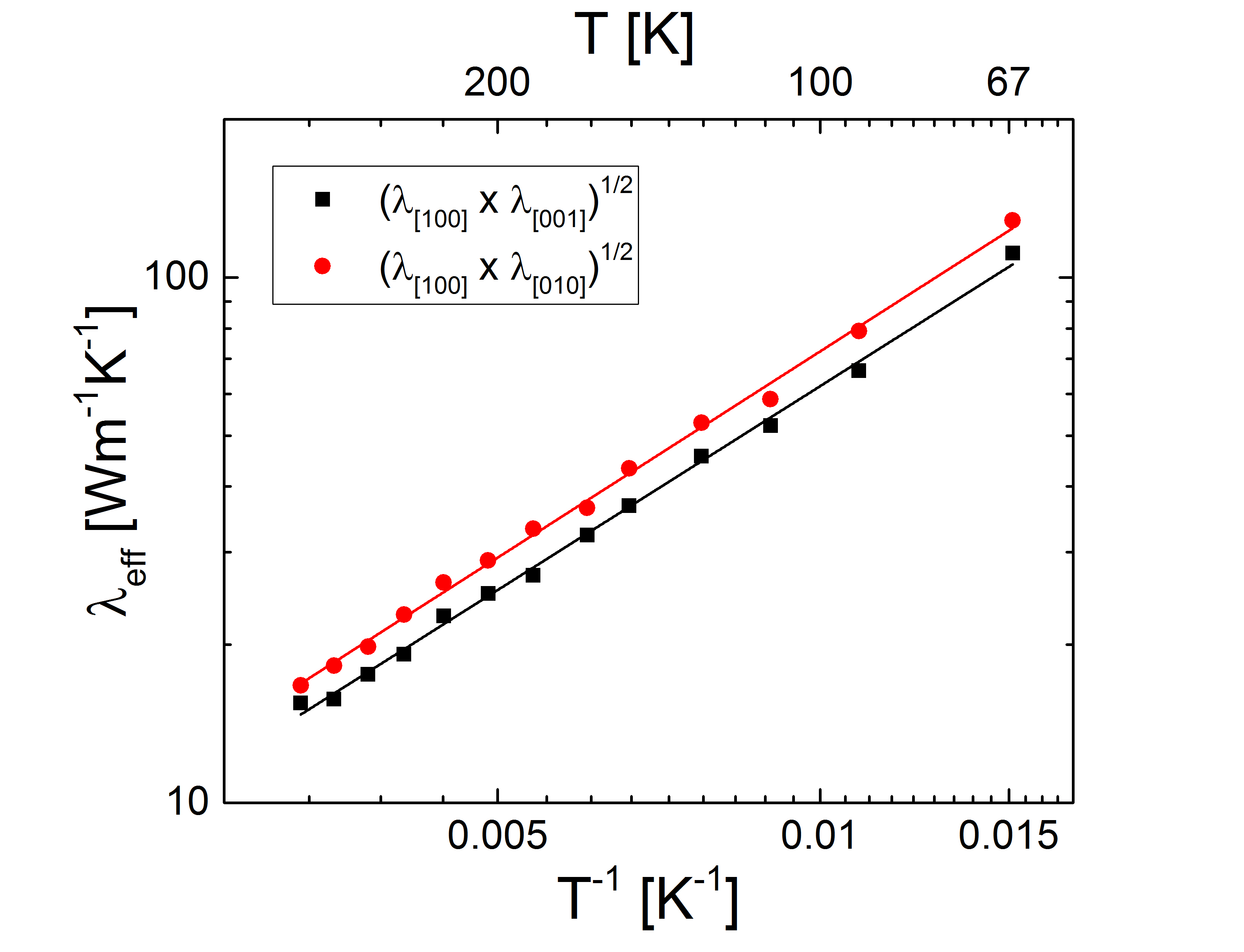}
\caption{Temperature dependent thermal conductivity of insulating $\beta$-$\mathrm{Ga_2O_3}$ on a logarithmic plot with a $\lambda(T)=b\cdot T^n$ approximation.}
\label{l}
\end{figure}

From the experiment we determine the temperature dependent thermal conductivity effective values $\lambda_\mathrm{ac}=\sqrt{\lambda_{[100]}\cdot\lambda_{[001]}}$ and $\lambda_\mathrm{ab}=\sqrt{\lambda_{[100]}\cdot\lambda_{[010]}}$ as shown in figure \ref{l}.
The anisotropy factor of the two effective thermal conductivity values squared should be the same as the anisotopy factor of the thermal diffusivities because
\begin{equation}
\left(\frac{\lambda_\mathrm{ab}}{\lambda_\mathrm{ac}}\right)^2=\left(\frac{\sqrt{\lambda_{[100]}\cdot\lambda_{[010]}}}{\sqrt{\lambda_{[100]}\cdot\lambda_{[001]}}}\right)^2=\frac{\lambda_{[010]}}{\lambda_{[001]}}=1.4\pm 0.1
\end{equation} 

The temperature depedence of $\lambda_\mathrm{ab}$ and $\lambda_\mathrm{ac}$ can be seen in figure \ref{l} approximated by the equation $\lambda(T)=b\cdot T^n$ with the slope $n=1.3\pm0.1$. 
For high temperatures $T>200~\mathrm{K}$ this slope value $n$ is found both experimentally by Guo \textit{et al.} \cite{Guo} and theoretically by Sanita \textit{et al.} \cite{Lambdatheo}. 
However, for low temperatures the dominance of phonon mean free path limiting scattering processes is dependent on the crystal quality and differs.
For room temperature using the specific heat $C_V=0.5~ \mathrm{kJkg^{-1}K^{-1}}=3~\mathrm{kgcm^{-3}K^{-1}}$ one can calculate all main axis thermal conductivity and diffusivity values as listed in table \ref{tabletop}.
\begin{table}
\begin{tabular}{ l | l | l | l | l | l }
axis & L.P. \cite{Struktur1} & $D$ & $\lambda$ & $\lambda_\mathrm{ex,ref}$\cite{Guo}& $\lambda_\mathrm{theo,ref}$\cite{Lambdatheo}\\
\hline
 & $\mathrm{\AA}$ & $\mathrm{mm^{2}s^{-1}}$ & $\mathrm{Wm^{-1}K^{-1}} $ & $\mathrm{Wm^{-1}K^{-1}} $ & $\mathrm{Wm^{-1}K^{-1}} $  \\
\hline\hline
\textit{a} [100] & 12.2 &  $3.7\pm 0.4$& $11\pm 1 $ &  $10.9\pm 1.0$ & $16$ \\
\textit{b} [010] & 3.0  & $9.6\pm 0.5$ & $29\pm 2 $ & $27.0\pm 2.0$ & $22$\\
\textit{c} [001] & 5.8  & $7.1\pm 0.4$ & $21\pm 2$ & $15$ & $21$  \\
\end{tabular}
\caption{Lattice parameter (L.P.), thermal diffusivity $D$ and thermal conductivity $\lambda$ values for the different crystal axis of our Czochralski grown $\beta$-$\mathrm{Ga_2O_3}$ and literature values $\lambda_\mathrm{ref}$ at room temperature.}
\label{tabletop}
\end{table}
The crystallographic order of the thermal conductivity values is in agreement with experimental work by Guo \textit{et al.} as well as the theoretical result of Sanita \textit{et al.} \cite{Lambdatheo}. 
The lowest thermal conductivity value is along the [100] axis, the highest for the [010] axis. 
The values of $\lambda_{[100]}$ and $\lambda_{[010]}$ are in accordance with Guo \textit{et al.} \cite{Guo} the value of $\lambda_{[001]}$ is in our case about $30\%$ higher than in that of Guo \textit{et al.}\cite{Guo}.
The anisotropy factors including the [100]-direction are ${\lambda_{[010]}}/{\lambda_{[100]}}=2.5\pm0.3$ and ${\lambda_{[001]}}/{\lambda_{[100]}}=1.9\pm0.2$. 
Additionally performed 3$\omega$ measurements confirm the presented values with a lower accuracy and are shown in the Appendix.\\

\section{Conclusion}

The temperature-dependent values of the thermal diffusivity in [010] and [001] direction as well as thermal conductivity in [100], [010] and [001] direction of Mg-doped insulating monoclinic $\beta$-$\mathrm{Ga_2O_3}$ bulk crystals were measured by the 2$\omega$-method. 
We find an anisotropic behavior for the thermal diffusivity and conductivity along [100], [010] and [001]. The anisotropy factor $D_{[010]}/D_{[001]}=\lambda_{[010]}/\lambda_{[001]}=1.4\pm 0.1$, ${\lambda_{[010]}}/{\lambda_{[100]}}=2.5\pm0.3$ and ${\lambda_{[001]}}/{\lambda_{[100]}}=1.9\pm0.2$ are temperature-independent. 
The room temperature values for the thermal conductivity in the main crystal axes are determined to $\lambda_\mathrm{[100]}=(11\pm 1)~ \mathrm{Wm^{-1}K^{-1}}$, $\lambda_\mathrm{[010]}=(29\pm 2)~ \mathrm{Wm^{-1}K^{-1}}$ and $\lambda_\mathrm{[001]}=(21\pm 2)~\mathrm{Wm^{-1}K^{-1}}$.
The temperature dependence of the thermal conductivity and diffusivity  confirms the model of phonon-phonon Umklapp scattering for $T>150 ~\mathrm{K}$. At lower temperatures the estimated thermal diffusivity and conductivity values are reduced by point-defect scattering.

\section*{Acknowledgments}

M.H. is grateful for financial support by 'MatSEC Graduate School', Helmholtz-Zentrum Berlin f{\"u}r Materialien und Energie GmbH.
Partial financial support by the German science foundation Fi932/7-2.
We gratefully acknowledge technical support from Hans Homburg and Jan Petrick.\\

\section*{Appendix}

\begin{figure}[h]
\includegraphics[width=0.7\columnwidth,keepaspectratio]{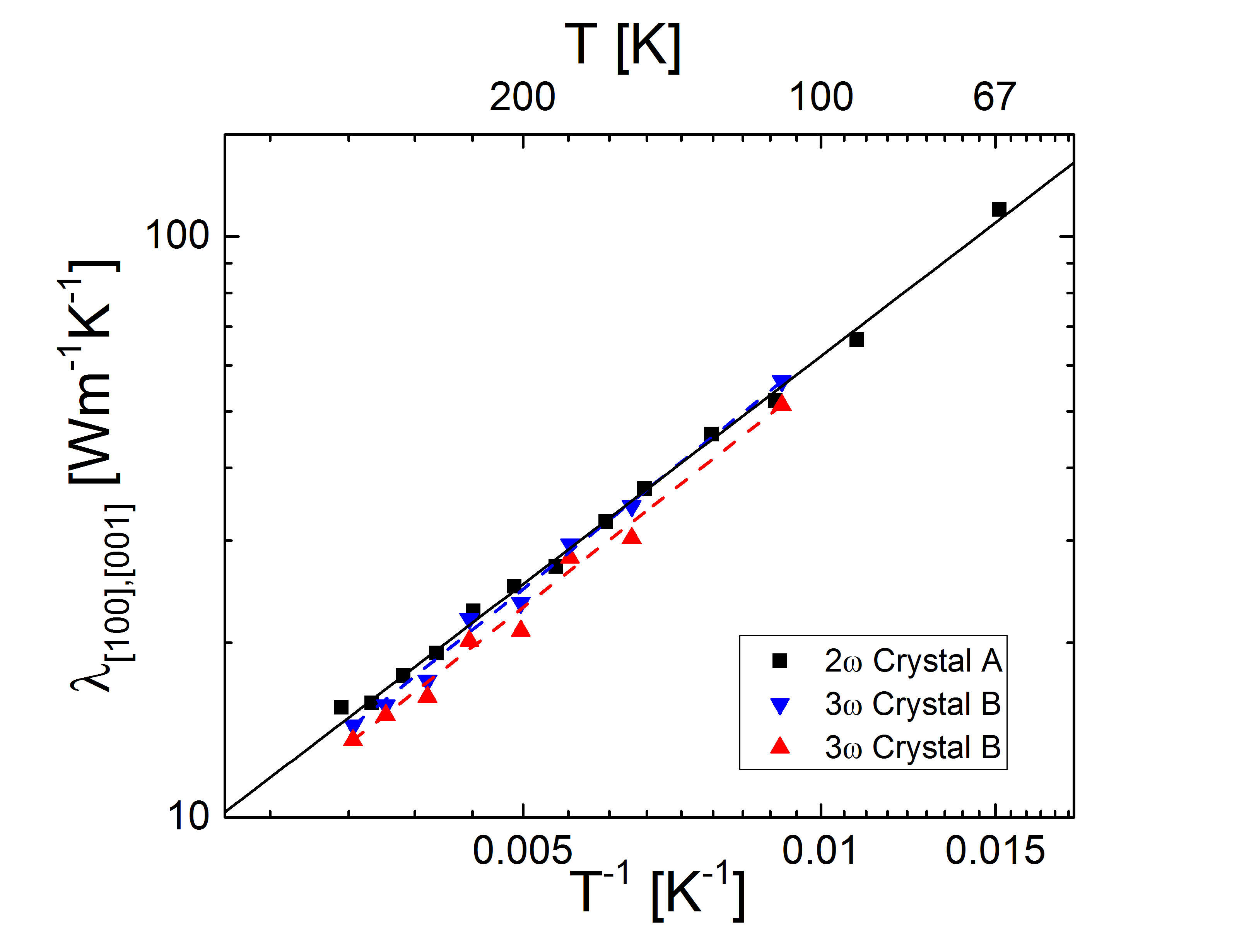}
\caption{Temperature dependent effective thermal conductivity values $\bar{\lambda}_\mathrm{[100],[001]}=\sqrt{\lambda_{[100]}\cdot\lambda_{[001]}}$ pictured in a logarithmic plot and approximated with $\bar{\lambda}=b\cdot T^n$. }
\label{lambda}
\end{figure}

Usually 3$\omega$ measurements are used to determine the thermal conductivity of materials. 
To confirm the presented values for the thermal conductivity one can compare the 2$\omega$-method used in this article with the 3$\omega$-method\cite{Cahill2} measurement results.
Here, we present measurements on a similar Mg-doped insulating $\beta$-$\mathrm{Ga_2O_3}$ single crystal with two heater lines to determine the effective thermal conductivity $\sqrt{\lambda_\mathrm{[100]}\times\lambda_\mathrm{[001]}}$ and compare it with the  values of the 2$\omega$-method shown in figure \ref{lambda}.

\begin{figure}
\includegraphics[width=0.7\columnwidth,keepaspectratio]{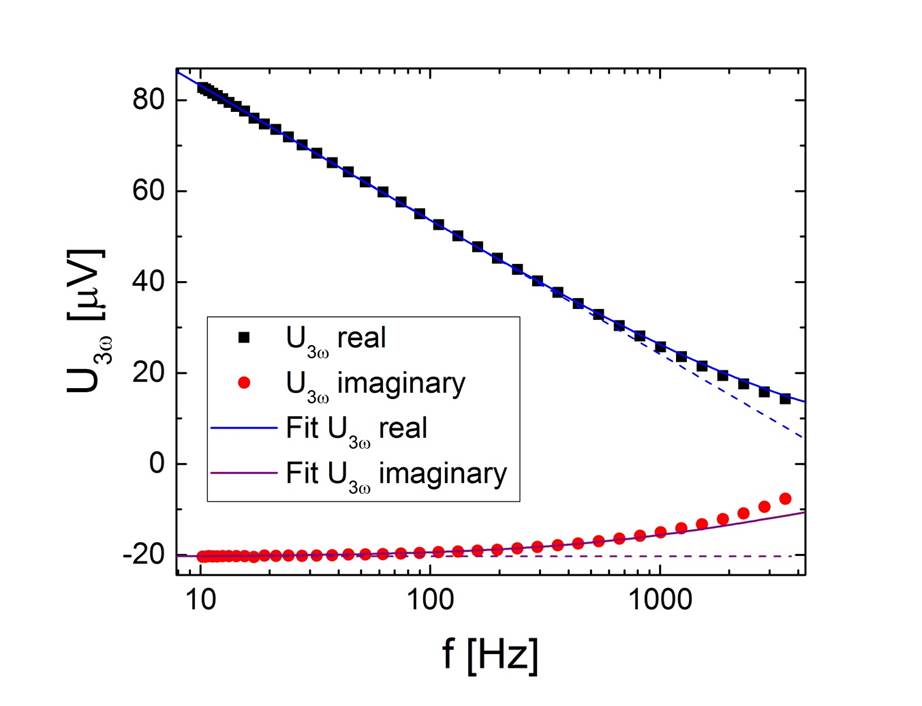}
\caption{Exemplary plot for the frequency-dependent voltage $U_{3\omega}(f)$ fitted with the approximation of equation \ref{3w}(solid lines).}
\label{u3wf}
\end{figure}

To determine the effective thermal conductivity with a single heater line one has to detect the temperature oscillations $\Delta T(2\omega)$ induced by the heater lines AC power $P(2\omega)$. 
These temperature oscillations lead to resistance oscillations in the heater line $R=R_0+R_0 \alpha\Delta T(2\omega)$.
The measured voltage $U$ contains two parts $U=R(2\omega)\cdot I(1\omega)=U(1\omega)+U(3\omega)$ with 
\begin{equation}
U(3\omega)=\frac{\alpha U_{1\omega}}{2}\Delta T\quad .
\end{equation}
To detect the rather small voltage $U(3\omega)$ within a huge voltage $U(1\omega)$ a Lock-In amplifier and a filter were used.
The frequency dependence of the temperature oscillations is
\begin{equation}
\Delta T(r)=\frac{P}{\pi l\bar{\lambda}}\int^\infty_0\frac{\sin(kw)}{kw\sqrt{k^2+q^2}}\mathrm{d}k \quad.
\label{3w}
\end{equation}
Here, we denote the inverse thermal penetration depth $q=\sqrt{4\pi f/D}$, the half heater line width $w$, the heating power $P$, the length of the heater line $L_\mathrm{heater}$.
Here, the effective thermal conductivity\cite{ramu2} is $\bar{\lambda}=\sqrt{\lambda_{[100]}\cdot\lambda_{[001]}}$.

The agreement of the measurement with this solution is shown in figure \ref{u3wf}.
The direct comparison of the obtained values of the $2\omega$ and the $3\omega$ measurements are shown in figure \ref{lambda}.

There is a good agreement between these two measurements with a better accuracy of the $2\omega$ measurement. 
The slopes $n$ of the approximation $\lambda=b\cdot T^n$ are $n_{2\omega}=1.3\pm 0.1$, $n_{3\omega,1}=1.4\pm 0.1$ and $n_{3\omega,2}=1.3\pm 0.1$ are the same. 
The room temperature ($T=297 \mathrm{K}$) values are $\bar{\lambda}_{2\omega}=15\pm 1$, $\bar{\lambda}_{3\omega,1}=14\pm 2$ and $\bar{\lambda}_{3\omega,2}=13\pm 2$.
There are three reasons why the $2\omega$ method is prefered to the $3\omega$ method.
First, we get well-defined thermal diffusivity values for specific directions.
Second, there are no parasitic $2\omega$ signals within the current, but every small heat source produces a $3\omega$ signal.
Third, the Lock-In amplifier does not have to take care of an oscillating $U_\mathrm{1\omega}$ signal.
\\
\section*{References}

\end{document}